\documentclass[preprint2]{aastex}

\newcommand{\rev}{}

\slugcomment{Submitted to ApJ}

\shorttitle{Angular Sizes of $\mu$Jy Radio Sources}
\shortauthors{Cotton et al.}

\begin{document}


\title{The Angular Size Distribution of $\mu$Jy Radio Sources}


\author{W.~D.~Cotton\altaffilmark{1}, J.~J.~Condon\altaffilmark{1},
K.~I.~Kellermann\altaffilmark{1}, M.~Lacy\altaffilmark{1},
R.~A.~Perley\altaffilmark{2}, A.~M.~Matthews\altaffilmark{3},
T.~Vernstrom\altaffilmark{4}, Douglas~Scott\altaffilmark{5}, J.~V.~Wall\altaffilmark{5}}
\affil{\altaffilmark{1}National Radio Astronomy Observatory, 520 Edgemont Road,
Charlottesville, VA 22903, USA} 
\affil{\altaffilmark{2}National Radio Astronomy Observatory, P.O. Box
0, Socorro, NM 87801, USA}
\affil{\altaffilmark{3}Astronomy Department, University of Virginia,
Charlottesville, VA, USA}
\affil{\altaffilmark{4}Dunlap Institute for Astronomy and
Astrophysics, University of Toronto, Toronto, ON M5S 3H4 Canada}
\affil{\altaffilmark{5}Department of Physics and Astronomy, University
of British Columbia, Vancover, BC V6T 1Z1, Canada}


\begin{abstract}

  We made two new sensitive (rms noise $\sigma_\mathrm{n} \approx\,1\,
  \mu$Jy beam$^{-1}$) high resolution ($\theta = 3\,\farcs0$ and
  $\theta = 0\,\farcs66$ FWHM) S--band ($2 < \nu < 4$~GHz) images
  covering a single JVLA primary beam ($\mathrm{FWHM} \approx
  14\arcmin$) centered on J2000 $\alpha =
  10^\mathrm{h}\,46^\mathrm{m}$, $\delta = +59^\circ\, 01\arcmin$ in
  the Lockman Hole.  These images yielded a catalog of 792 radio
  sources, $97.7 \pm 0.8$\% of which have infrared counterparts
  stronger than $S \approx 2\,\mu\mathrm{Jy}$ at $\lambda =
  4.5\,\mu\mathrm{m}$.  About 91\% of the radio sources found in our
  previously published, comparably sensitive low resolution ($\theta =
  8\arcsec$ FWHM) image 
  covering the same area were also detected at $0\,\farcs66$
  resolution, so most radio sources with $S(3\,\mathrm{GHz}) \gtrsim
  5~\mu\mathrm{Jy}$ have angular structure $\phi \lesssim
  0\,\farcs66$.  The ratios of peak { brightness} in the
  $0\,\farcs66$ and $3''$ images have a distribution indicating that
  most $\mu$Jy radio sources are quite compact, with a median Gaussian
  angular diameter $\langle \phi \rangle = 0\,\farcs3 \pm 0\,\farcs1$
  FWHM and an rms scatter $\sigma_\phi \lesssim 0\,\farcs3$ of
  individual sizes.  Most of our $\mu$Jy radio sources obey the tight
  far-infrared/radio correlation, indicating that they are powered by
  star formation.  The median effective angular radius enclosing half
  the light emitted by an exponential disk is $\langle \rho_\mathrm{e}
  \rangle \approx \langle \phi \rangle / 2.43 \approx 0\,\farcs12$, so
  the median effective radius of star-forming galaxies at redshifts $z
  \sim 1$ is $\langle r_\mathrm{e}\rangle \approx 1.0\mathrm{~kpc}$.

\end{abstract}



\section{Introduction}
We recently reported the results of a low-resolution ($\theta =
8\arcsec$ FWHM) S-band ($2 < \nu < 4$~GHz) image
made with the
NRAO\footnote{The National Radio Astronomy Observatory is a facility
of the National Science   Foundation operated under cooperative
agreement by Associated  Universities, Inc.} Karl G. Jansky Very Large
Array (VLA) C 
configuration, covering a single primary beam (FWHM~$\approx14\arcmin$)
centered on J2000 $\alpha = 10^\mathrm{h}\, 46^\mathrm{m}$, $\delta =
+59^\circ \,01\arcmin$ in the Lockman Hole \citep{Paper1,Paper2}.
The rms noise and confusion in this image are comparable: $\sigma_\mathrm{n}
\approx \sigma_\mathrm{c} \approx 1~\mu \mathrm{Jy~beam}^{-1}$.  The
rapidly falling Euclidean-normalized differential source count
$S^{5/2}n(S)$ at $\mu$Jy levels obtained from the confusion amplitude or
``$P(D)$'' distribution,  converted to 1.4~GHz via the median
spectral index $\langle \alpha \rangle = -0.7$, closely follows
predictions of evolutionary models \citep{Condon84,Wilman2008} in
which most $\mu$Jy radio sources are powered by recent star formation
in galaxies at median redshift $\langle z \rangle \sim 1$.

Our $\mu$Jy source count is much lower than that of \citet{OM2008},
who derived a nearly constant $S^{5/2}n(S)$ from their sensitive 1.4~GHz
VLA image made at the same position with $\theta = 1 \farcs 6$
resolution.  
\cite{OM2008} corrected their source count for the
effects of partial resolution using a source angular-size distribution
with median FWHM $\langle \phi \rangle \approx 1 \farcs2$ and a tail
extending to much larger angular sizes.  
\cite{Paper1} detected all of the overlapping \citet{OM2008} sources,
so it appears that the count correction required by the broad source
angular-size distribution used is the primary cause of the higher
\cite{OM2008} source count.   
A steep differential source count $n(S) \propto S^{-5/2}$
magnifies the impact of overestimated angular sizes on count
corrections.  A similar effect has appeared in numerous published
source counts at {\rev sub-}mJy levels, all of which were corrected for a range of
assumed angular-size distributions and consequently disagree by
amounts far greater than the published uncertainties
\citep{con07,Heywood2013}.

F.~Owen (private communication) recently made a sensitive VLA 1.5~GHz
image ($\theta = 1\farcs6$ resolution, $\sigma =
2.2\,\mu\mathrm{Jy~beam}^{-1}$ {\rev rms} noise) of the GOODS-N field and
found a typical FWHM source size $\langle \phi \rangle \sim 1\arcsec$.
However, this result depends on data from a single VLA configuration
and does not include multi-{\rev configuration} images made with different VLA
configurations {\rev and include tests using simulated data }
(of the type presented in Section \ref{simulations}) to
demonstrate {\rev that} the techniques {\rev used} can robustly 
recover source sizes near or slightly below the instrumental resolution.

A sensitive
high-resolution ($\theta \approx 0\,\farcs22$) survey of the GOODS-N
field using the VLA at 10 GHz \citep{Murphy2017} recently detected a
sample of 32 sources with a much smaller median angular diameter
$\langle \phi \rangle = 0\,\farcs167 \pm 0\,\farcs032$ and rms size
scatter $\sigma_\phi \approx 0\,\farcs091$.  Radio sources in star-forming galaxies are
expected to be smaller at 10 GHz, owing to the stronger contribution of
free-free emission, while the synchrotron radiation dominating at lower
frequencies is spread out by cosmic-ray diffusion.  However, the
cosmic-ray diffusion is too small to grow sources from $0\,\farcs17$ at 10~GHz
to  $1\arcsec$ at 1.5~GHz.  Thus, there is still a significant spread
in the reported median angular diameters of $\mu$Jy radio sources, and
they remain disturbingly correlated with the resolution of the images
used to find and fit Gaussians to the sources.

We believe that previous attempts to measure angular-size
distributions and counts of faint radio sources have disagreed largely
because: (1) they depended on sensitive images made with only a single
high-resolution antenna configuration; (2) high-resolution images miss
extended radio emission whose surface brightness is below the
detection limit at that resolution; and (3) image noise tends to
broaden Gaussians fitted to faint point sources by amounts
proportional to the image resolution $\theta$.  A more reliable method
for constraining the angular-size distributions of source populations
is to measure the peak flux densities (so-called ``peak flux
densities'' are actually specific intensities written in units of flux
density per beam solid angle; e.g., $\mu$Jy beam$^{-1}$) in two or
more images made with different array configurations, yielding
different angular resolutions but similar point-source sensitivities.
See Appendices B and C of \citet{Murphy2017} for a discussion of
this method.

This paper presents two new S-band images made with $\theta =
3\,\farcs0$ and $\theta = 0\,\farcs66$ resolution from VLA B- and
A-configuration data, respectively.  Both are centered on J2000
$\alpha = 10^\mathrm{h} \, 46^\mathrm{m}$, $\delta = +59^\circ
\,01\arcmin$.  Extensive optical and infrared (IR) material on this
field is available from \citet{Strazzullo2010}, \citet{Mauduit2012},
and \citet{Herschel160}.  Both images have $\sigma_\mathrm{n} \approx
1~\mu\mathrm{Jy~beam}^{-1}$ rms noise and negligible
confusion. The new $3\arcsec$ and earlier $8\arcsec$ images yield
accurate total flux densities because they do not resolve the vast
majority of the $\mu$Jy source population, while the $\theta =
0\,\farcs66$ image should marginally resolve the $\phi\lesssim
0\,\farcs5$ sources expected in the star-forming galaxies at redshifts
$z \sim 1$ that dominate the $\mu$Jy source population
\citep{Condon84,Wilman2008}.

An initial analysis of the source population found in our $8''$
resolution image was given in \citet{Paper3}, who reported 10\%
radio-loud active galactic nuclei (AGNs), 28\% radio-quiet AGNs, 58\%
star--forming galaxies, and 4\% that could not be classified.
\citet{Paper3} also gave a traditional discrete source count that
needed little correction for partial resolution and agrees well with
the deeper $P(D)$ analysis.

\section{Observations\label{obs}}
The dates and durations of our sensitive VLA S-band observations centered on J2000
$\alpha=10^\mathrm{h} \,46^\mathrm{m}$, $\delta=+59^\circ \,01\arcmin$
are summarized in Table~\ref{ObsTab}.  The
C-configuration observations were described in \citet{Paper1},
\citet{Paper2}, and \citet{Paper3}.  The ``B+'' data were taken in the
BnA configuration and during the transition to A configuration, but
they had insufficient temporal or frequency resolution for the longer
BnA baselines to be used in our highest-resolution ($\theta =
0\,\farcs66$) A-configuration image.

\begin{table}
\caption{Observations.}\label{ObsTab}
\vskip 0.1in
\begin{center}
  \begin{tabular}{lccr}
\tableline
Array & Date Range & Time & No.  \\
\tableline    
 C & 2012 Feb 21 -- Mar 18  & 57.0  & 6  \\
 B+& 2014 Feb 02 -- Feb 18  & 26.0  & 10 \\
 A & 2015 Jul 10 -- Sep 13 & 39.4  & 6  \\
\hline
  \end{tabular}
  \end{center}
  \hfill\break

``Array'' is the VLA configuration. The start and end dates give the
period over which the data were taken. ``Time'' is the total observing
time in hours, and ``No.'' is the number of separate observing
sessions.
The ``C'' observations used the array prior to fully outfitting with 
3~GHz receivers and included only 21 antennas.

\end{table}

The new A-configuration data have higher frequency resolution
(500~kHz) and were recorded with shorter (1\,s) basic integration
times to minimize
bandwidth- and time-smearing within the half-power circle of the VLA
primary beam. The VLA correlator also separated the observed frequency
range $1.989 \leq \nu~\mathrm{(GHz)} \leq 4.013$ into 16 contiguous
subbands of width $\Delta \nu = 128$~MHz each.  The A-configuration
data processing was similar to that described in \citet{Paper1}. Our
flux-density calibration is based on a standard spectrum and model of
3C147 \citep{PB2013A} and was transferred to the unresolved phase
calibrator J1035+5628, whose absolute position uncertainty is $<
0\,\farcs002$. We used J1035+5628 to determine instrumental
polarization and 3C147 \citep{PB2013B} to calibrate the
cross-polarized delays and phases.  Calibration and editing used
standard scripts in the Obit package \citep{Obit}\footnote{\rev Obit
software and documentation are available from  
\url{http://www.cv.nrao.edu/$\sim$bcotton/Obit.html}.}. 
After calibration
and extensive editing, the data were averaged over baseline-dependent
time intervals chosen to minimize the size of the $(u,v)$ data set but
avoid time smearing.

\section{Imaging}
We imaged the data sets using a joint multi-frequency CLEAN that both
minimizes frequency dependent effects and exploits the full
sensitivity of the wideband data.  A single-resolution CLEAN was
adequate for this field dominated by nearly unresolved sources.
{\rev The low resoution image was imaged and restored using a
$3\arcsec$ FWHM round, Gaussian beam and the high resolution data with a
$0\,\farcs66$ FWHM round, Gaussian beam.
These values correspond to the resolution near the bottom of the ($2 <
\nu < 4$~GHz) band.
Briggs' ``optimal robust'' weighting was used in the image formation
\citep{Briggs}. 
}

\subsection{Wide-band, Wide-field Imaging}
The large fractional bandwidth and wide field-of-view {\rev that was}
imaged require 
that both the source spectra and the antenna gain as a function of
position and frequency be taken into account.  This was done by the
Obit task MFImage, which divides the observed spectrum into frequency
bins narrow enough that the variations in antenna gain and spectral
differences among sources are small within each bin.  For this
purpose, we set the frequency bin width equal to the 128~MHz
correlator subband width.

The image was divided into a large number of small facet planes to
minimize the effects of sky curvature.  The facets were reprojected
onto a common tangent plane and grid to allow parallel CLEANing.  A
frequency-dependent $(u,v)$ taper was applied to keep the angular
resolution of the dirty beam the same in all frequency bins.  This,
plus the use of a single restoring beam, yields a meaningful
16-channel spectrum in each spatial pixel of the image cube.

For each major cycle of CLEAN, dirty and residual images were computed
separately for each of the 16 frequency bins.  The more sensitive
full-bandwidth image, derived from the $\sigma_i^{-2}$ noise-weighted
average of the frequency-bin images, was used to drive the minor cycle
CLEANing.  The sensitive combined image and the combined dirty beam
were used to locate new CLEAN components, and the dirty beam for the
corresponding frequency bin and facet was used to derive the residuals
for the next minor cycle.

Once the minor cycles hit their stopping criteria, the accumulated CLEAN
model was subtracted from the visibility data.
The CLEAN model subtracted from each frequency bin used the CLEAN flux
density of each component in that bin corrected in frequency by
the spectral index fitted to each component using all frequency bins.
This process was accelerated by a Graphics Processing Unit (GPU).

{\rev After} the CLEANing was done, the image in each frequency bin was
restored with the components subtracted from that bin convolved with the
single Gaussian restoring beam fitted to the central facet
of the combined full-sensitivity image.

This procedure accommodates variations with frequency of antenna gain
and source flux density by using frequency bins sufficiently narrow
that variations within a bin do not disturb the image quality.  The
spectrum in a given pixel depends on both antenna gain and source
spectral index.  The antenna gain as a function of position and
frequency was measured independently \citep{Perley16}, allowing the 
spectral indices of sufficiently strong sources to be determined.  For
weaker sources, we used the average spectral index $\langle \alpha
\rangle = -0.7$ to fit the source flux density at any frequency.
That approximation is valid for most $\mu$Jy sources at S band.

Faraday rotation in the Stokes {\it Q} and {\it
  U} images is preserved if the frequency bins are sufficiently narrow
and the rotation measure is not too large.  A rotation measure
$\mathrm{RM} = \pm 2000 \mathrm{~rad~m}^{-2}$  rotates the
polarization position angle by $1/2$ turn across a 128~MHz frequency
bin at $\nu \approx 3\,\mathrm{GHz}$, so larger RMs {\rev than this}
will cause significant Faraday depolarization.

  \subsection{CLEAN Windows}
  
CLEAN deconvolution works best if it is constrained to place
components only in spatial ``windows'' containing actual emission.  Our CLEAN
windows were generated or updated at the beginning of each major cycle
by the combined (wideband) image covering each facet. If the maximum
residual lay outside the current window and its peak was higher than
5 times the facet rms, a new round window was added to the
existing CLEAN window at the location of the peak, with a radius
derived from the structure function about the peak.  This allows
CLEANing down to (or into) the noise and captures the bulk of the
emission in the CLEAN model without producing excessive CLEAN bias.

\subsection{Self-calibration}

A single phase self-calibration was applied to the data.  The model
visibilities were calculated as was done in the CLEAN major cycles,
and an independent phase solution was determined for each 10 minutes
in each spectral window and polarization.  These
phases were interpolated in time and {\rev applied to} all data.

\subsection{Image Adjustments}

The resulting images are 16 spectral-channel
cubes covering $2 < \nu\mathrm{(GHz)} < 4$
that were jointly deconvolved and restored with a common spatial
resolution.  The sky {\rev brightness} $S_\mathrm{p}(x,y)$ of the pixel
offset by $(x,y)$ from the pointing center at the
reference frequency $\nu_0 = 3$~GHz was calculated from the noise-weighted
average of the spectral-window images:
\begin{equation}\label{eqn:refflux}
S_\mathrm{p}(x,y)=  \frac {\sum\limits_{i=1}^{16} {{S_i(x,y) \exp[{-\alpha
    {\rm ln}({\nu_i /{\nu_0}})}]}\over{g_i(x,y)}} 
{g^2_i(x,y)\over{\sigma^2_i}} }
{ \sum\limits_{i=1}^{16} {g^2_i(x,y)\over{\sigma^2_i}}}~,
\end{equation}
where $S_i(x,y)$ is the image (not corrected for primary-beam
attenuation) {\rev brightness} at pixel position $(x,y)$ in spectral
channel $i$, $\alpha \approx -0.7$ is the median source spectral
index, $\nu_i$ is the central frequency of spectral channel $i$, $0
\leq g_i(x,y) \leq 1$ is the normalized antenna gain at offset
$(x,y)$ from the pointing center and frequency $\nu_i$, and $\sigma^2_i$ is the mean variance
of source-free regions in the $i$th spectral image.  The first factor
in the numerator of Equation~\ref{eqn:refflux} corrects the bin {\rev brightness}
$S_i(x,y)$ to the reference frequency using spectral
index $\alpha$ and divides it by the primary attenuation $g_i(x,y)$ to
yield the {\rev brightness} on the sky $S_\mathrm{p}(x,y)$.  The
second factor is the pixel  weight that maximizes the signal-to-noise
ratio, specifically the antenna gain $g_i^2(x,y)$ divided by the
spectral-channel image variance $\sigma_i^2$.

We approximated the normalized VLA antenna power gain $g_i(x,y)$ by
the theoretical gain of a uniformly illuminated circular aperture:
\begin{equation}\label{eqn:jincsq}
  g_i(\rho) =  \mathrm{jinc^2}(D\rho/\lambda_i) 
  \equiv \Biggl[ \frac {2 J_1(\pi D \rho /\lambda_i)}{\pi D \rho /
  \lambda_i} \biggr]^2~,
\end{equation}
where $\rho = (x^2 + y^2)^{1/2}$ is the offset from the pointing
center in radians, $J_1$ is the Bessel function of the first kind of
order 1 \citep{Bracewell}, $D = 25\,\mathrm{m}$ is the aperture
diameter, and $\lambda_i = c / \nu_i$ is the wavelength at the center
frequency $\nu_i$ of spectral channel $i$.  Equation~\ref{eqn:jincsq}
yields the primary beam FWHM $\theta_{1/2}$ in practical units; it is
\begin{equation}
  \biggl( \frac{\theta_{1/2}}{\mathrm{arcmin}} \biggr) \approx
  42.0 \biggl( \frac{\mathrm{GHz}}{\nu} \biggr)~,
\end{equation}
which is only $\sim1$\% wider than the average beamwidth measured
across S band \citep{Perley16}.

Applying Equation~\ref{eqn:refflux} corrects source flux densities for
antenna gain but causes the image noise to increase radially as
$g_0^{-1} (\rho)$ away from the pointing center.  In order to
determine meaningful noise statistics in the neighborhood of each
source, we multiplied the gain-corrected image by the antenna gain
$g_0(x,y)$ at the reference frequency $\nu_0 = 3$~GHz during source
finding and fitting.  For example, our $0\,\farcs66$ resolution
A-configuration image extends to a radius $\rho = 8\,\farcm3$.  After
multiplication by $g_0(\rho)$, its mean off-source rms is $\sigma =
0.96~\mu$Jy beam$^{-1}$.

We made a comparable image with $3\arcsec$ resolution and radius $\rho =
13\,\farcm8$ using {\rev the combined uv}data from the longer
C-configuration baselines and the shorter BnA-configuration baselines
(Table~\ref{ObsTab}).   
Additionally, we convolved the A-configuration image to $3\arcsec$ resolution.
Differences in celestial position and flux density scale of the two
images were determined from the {\rev ``at most''} marginally resolved sources
brighter than 100~$\mu$Jy beam$^{-1}$.
Finally, the B+C-configuration $3\arcsec$ image was shifted in position
($0\farcs006$ in $\alpha$ and $-0\farcs079$ in $\delta$) and
scaled in flux density (0.933) to agree with the $3\arcsec$
A-configuration smoothed image,
and the two $3\arcsec$ images were weighted by $\sigma^{-2}$ and
combined.  
{\rev The differences in calibration are likely due to the extended
period over which the B+C data were taken and the difficulties of
calibration in a fierce RFI environment.
The {\rev rms} of the B+C image {\rev prior to primary beam
correction} was 1.27~$\mu$Jy beam$^{-1}$ and for the  
A-configuration image convolved to $3\arcsec$ was 1.88 $\mu$Jy beam$^{-1}$,
giving relative weights of 70\% and 30\% respectively.
}
Source-free regions in the final {\rev combined} $3\arcsec$ image have
rms $\sigma =1.01~\mu$Jy beam$^{-1}$ {\rev prior to primary beam
correction}.

The $3\arcsec$ resolution image has imaging artifacts near the strongest source in
the field, a hot spot in the lobe of an FR II  source.  To avoid
contaminating the image statistics, we {\rev masked these artifacts} by
assigning affected pixels a value indicating that they should be
ignored in subsequent analysis.


\section{Radio Source Catalog\label{Catalog}}
The Obit task FndSou was used to generate two independent lists of
radio components from the wideband $\theta = 3''$ and $\theta =
0\,\farcs66$ images, uncorrected for primary beam attenuation.
FndSou locates ``islands'' of contiguous pixels brighter than a chosen
peak flux density threshold and fits one or more elliptical Gaussian
components to the emission in each island.  These fits are subject to
a number of constraints; in particular, any fitted Gaussian narrower
than the (circular) CLEAN restoring beam is unphysical and was
fitted by a Gaussian at least as wide as the restoring beam.

FndSou initially fits Gaussians in one island at a time and ignores
overlapping components in adjacent islands.  After the first component
list for each image was generated, the parameters of potentially
overlapping components were reconciled by refitting them jointly with
all other components lying within 25 pixels ($2\,\farcs3$ on the
$\theta = 0\,\farcs66$ image or $11\farcs0$ on the $\theta = 3''$
image) in both $\alpha$ and $\delta$.  The sky peak flux density
$S_\mathrm{p}$ {(\rev $S_\mathrm{p}^\mathrm{L}$  for $\theta = 3''$
and $S_\mathrm{p}^\mathrm{H}$ for $\theta = 0\,\farcs66$})
of each Gaussian component was obtained by
interpolating between image pixels to its fitted centroid position
and dividing by the 3~GHz primary attenuation $g_0$ at that position.
{\rev Interpolations used the Lagrangian technique with a $5\times5$ kernel.}
Resolved sources represented by multiple components were replaced by
only the component closest to the IR galaxy position; these six sources
are discussed {\rev further} in Section \ref{extended}.

The component lists from the $\theta = 0\,\farcs66$ and $\theta = 3''$
images were merged to form a single list of source
candidates lying inside the circle with 3~GHz primary attenuation $g_0
> 0.25$ {\rev(radius $10\farcm1$)}.  The candidate list includes:
\begin{enumerate}
\item all components from the $\theta = 3''$ image with 
local signal-to-noise ratio $\mathrm{SNR} \equiv S_\mathrm{p}^\mathrm{L} / \sigma \geq 4$, plus
\item a small number of additional components with $\mathrm{SNR}
\equiv S_\mathrm{p}^\mathrm{H} / \sigma \geq 5$ from the $\theta = 0\,\farcs66$  image.
\end{enumerate}

For each of the candidate sources (item 1 above) the $\theta =
0\,\farcs66$ catalog was searched for nearby components with
$S_\mathrm{p}^\mathrm{H} \geq 3\,\mu\mathrm{Jy~beam}^{-1} \approx 3 \sigma$.
An IR counterpart of each of the candidate sources was sought as described
in Section \ref{IRId}.

From the candidate list,  we kept as source components only:
\begin{enumerate}
\item the complete sample of  596 candidates with $\mathrm{SNR} \geq 5$ on the
  $\theta = 3''$ image, plus 
\item the reliable but incomplete sample of 156 candidates with $4 \leq
  \mathrm{SNR} < 5$ from the $\theta = 3''$ image that were confirmed by an
   $S_\mathrm{p}^\mathrm{H} \gtrsim 3 \sigma$ component lying
  within $3''$ on the  $\theta = 0\,\farcs66$ image.
{\rev (Excluding regions blocked by the bright IR sources, 88\% of these
also were within ${\rev 1\farcs5}$ of an IR source.) }
Plus 
\item  
the additional 40 candidates with $4 \leq \mathrm{SNR} < 5$ from the $\theta =
3''$ image that were confirmed only by an IR source lying within
${\rev 1\farcs5}$ {\rev( see below)}.
  \end{enumerate}
Our final catalog contains 792 radio source components, a sample of
which is shown in Table \ref{ArcadeACatalog}; the full table is
available online.  At $\mu$Jy levels there are very few resolved
double radio sources, so nearly every cataloged radio source component
is also a complete astrophysical radio source, defined as all of the
radio emission from a single galaxy or AGN.  Of the 209 source
candidates from the $\theta = 3''$ image with $4 \leq \mathrm{SNR} <
5$ 
{\rev and in locations where an IR identification was possible}, only
14 had neither an IR identification nor a radio counterpart on the
$\theta = 0\,\farcs66$ image.
{\rev 75\% of these (154/205) were within ${\rev 1\,\farcs5}$ of an IR source and
outside of areas blocked by the brighter IR sources (see below).}

\begin{deluxetable}{ccccccccccc}
\tabletypesize{\small}
\rotate
\tablecaption{Source Catalog.\label{ArcadeACatalog}}
\tablewidth{0pt}
\tablehead{
  \colhead{J2000 $\alpha$} & \colhead{J2000 $\delta$} & \colhead{$\theta$} & \colhead{$g_0$} &
  \colhead{ $S_\mathrm{p}^\mathrm{L}$ } & \colhead{ $~S^\mathrm{L}$ } & \colhead{ L$-$H } & \colhead{ $S_\mathrm{p}^\mathrm{H}$ } & \colhead{ $~S^\mathrm{H}$ } & 
  \colhead{$\phi$} & \colhead{ $r$ } \\
  \colhead{{\llap h}~~m~~s~~~~~~~~~~~~s~}   & \colhead{$^\circ$~~$^\mathrm{m}$~~$''$~~~~~~~~~~$''$}  & \colhead{} & \colhead{} &
  \colhead{($\mu$Jy~beam$^{-1}${\rlap )}} & \colhead{~($\mu$Jy)}  &  \colhead{($''$)} & \colhead{($\mu$Jy~beam$^{-1}${\rlap )}} & \colhead{~($\mu$Jy)}  &
  \colhead{($''$)} & \colhead{($''$)} \\ 
}
\startdata
10 44 43.071 $\pm$ 0.071 & 58 59 51.05 $\pm$ 0.55 & L & 0.26 &   19.49 $\pm$  3.98 &  25.70 $\pm$  5.76 &    4.53 &                     &                     &           &      1.10  \\
10 44 44.109 $\pm$ 0.009 & 59 00 19.36 $\pm$ 0.07 & H & 0.27 &   16.96 $\pm$  4.02 &  16.00 $\pm$  3.82 &    0.20 &   21.18 $\pm$  3.31 &   20.66 $\pm$  3.29 &  $ <0.61$ &      0.24  \\
10 44 46.577 $\pm$ 0.015 & 58 58 40.70 $\pm$ 0.11 & H & 0.28 &   14.96 $\pm$  3.66 &  14.06 $\pm$  3.47 &    0.35 &   12.07 $\pm$  3.30 &   11.17 $\pm$  3.07 &  $ <0.86$ &      0.24  \\
10 44 46.940 $\pm$ 0.007 & 59 01 56.37 $\pm$ 0.04 & H & 0.30 &   39.90 $\pm$  3.46 &  44.51 $\pm$  4.26 &    0.13 &   33.50 $\pm$  3.01 &   33.23 $\pm$  3.15 &  $ <0.45$ &      0.50  \\
10 44 47.417 $\pm$ 0.013 & 58 58 01.79 $\pm$ 0.10 & H & 0.27 &   18.33 $\pm$  3.76 &  17.56 $\pm$  3.64 &    0.23 &   13.59 $\pm$  3.25 &   12.82 $\pm$  3.09 &  $ <0.79$ &      0.18  \\
10 44 47.563 $\pm$ 0.006 & 58 59 18.74 $\pm$ 0.03 & H & 0.30 &   49.70 $\pm$  3.60 &  49.44 $\pm$  3.88 &    0.13 &   55.51 $\pm$  3.04 &   55.34 $\pm$  3.45 &  $ <0.36$ &      0.29  \\
10 44 47.661 $\pm$ 0.005 & 59 00 35.66 $\pm$ 0.03 & H & 0.31 &   88.76 $\pm$  3.61 &  91.48 $\pm$  4.67 &    0.06 &   79.23 $\pm$  3.02 &   80.61 $\pm$  3.93 &  $ <0.33$ &      0.38  \\
10 44 47.725 $\pm$ 0.018 & 59 02 14.60 $\pm$ 0.14 & H & 0.31 &   34.85 $\pm$  3.39 &  35.53 $\pm$  3.66 &    0.82 &   11.34 $\pm$  3.04 &   12.69 $\pm$  3.66 &  $ <1.15$ &      0.83  \\
10 44 47.794 $\pm$ 0.012 & 58 58 12.47 $\pm$ 0.08 & H & 0.28 &   19.53 $\pm$  3.69 &  19.97 $\pm$  3.93 &    0.54 &   15.82 $\pm$  3.22 &   15.16 $\pm$  3.12 &  $ <0.72$ &      0.77  \\
10 44 49.188 $\pm$ 0.014 & 58 57 19.82 $\pm$ 0.10 & H & 0.27 &   17.85 $\pm$  3.82 &  17.04 $\pm$  3.68 &    0.65 &   12.71 $\pm$  3.28 &   11.86 $\pm$  3.08 &  $ <0.83$ &      0.47  \\
\enddata

\tablecomments{Table~\ref{ArcadeACatalog} is published in its entirety
 in machine-readable format.  A portion is shown here for guidance
 regarding its form and content. The table lists J2000 right
 ascensions $\alpha$ and declinations $\delta$ measured from the
 $\theta = 0\,\farcs66$ resolution image if available (indicated by ``H" in
 the $\theta$ column), otherwise from the $\theta = 3''$ resolution image
 (``L'' in the $\theta$ column).  
 {\rev Interpretation of the fitted Gaussian {\rev parameters} follows the
 development in \citep{con97}.}
 The rms position errors include our
  estimated absolute astrometric uncertainty $\sigma_\alpha =
  \sigma_\delta \approx 0\,\farcs02$. Column $g_0$ gives the
  normalized antenna gain at the source position.  The 3~GHz peak and
  total flux densities corrected for fitting bias ({\rev
  \cite{con97}}) from the 
  low-resolution image are listed under $S_\mathrm{p}^\mathrm{L}$ and
  $S^\mathrm{L}$.  L$-$H is the separation of the positions measured
  on the low- and high-resolution images.  $S_\mathrm{p}^\mathrm{H}$
  and $S^\mathrm{H}$ are the peak and integrated flux densities from
  the high-resolution image.  Column $\phi$ gives the deconvolved
  Gaussian source FWHM sizes or upper limits at $0\,\farcs66$
  resolution. {\rev Next,} $r$ is the angular distance between the radio source and
  its nearest IR neighbor.  Separations less than $1\,\farcs5$ are
  considered solid associations, $1\,\farcs5$ $\leq r \leq$ $3\,\farcs5$ are
  probable associations,  and $r > 3\,\farcs5$ are unassociated.}
\end{deluxetable}

\section{Radio/IR Identifications\label{IRId}}

\begin{figure*}
\centerline{
  \includegraphics[width=7.0in]{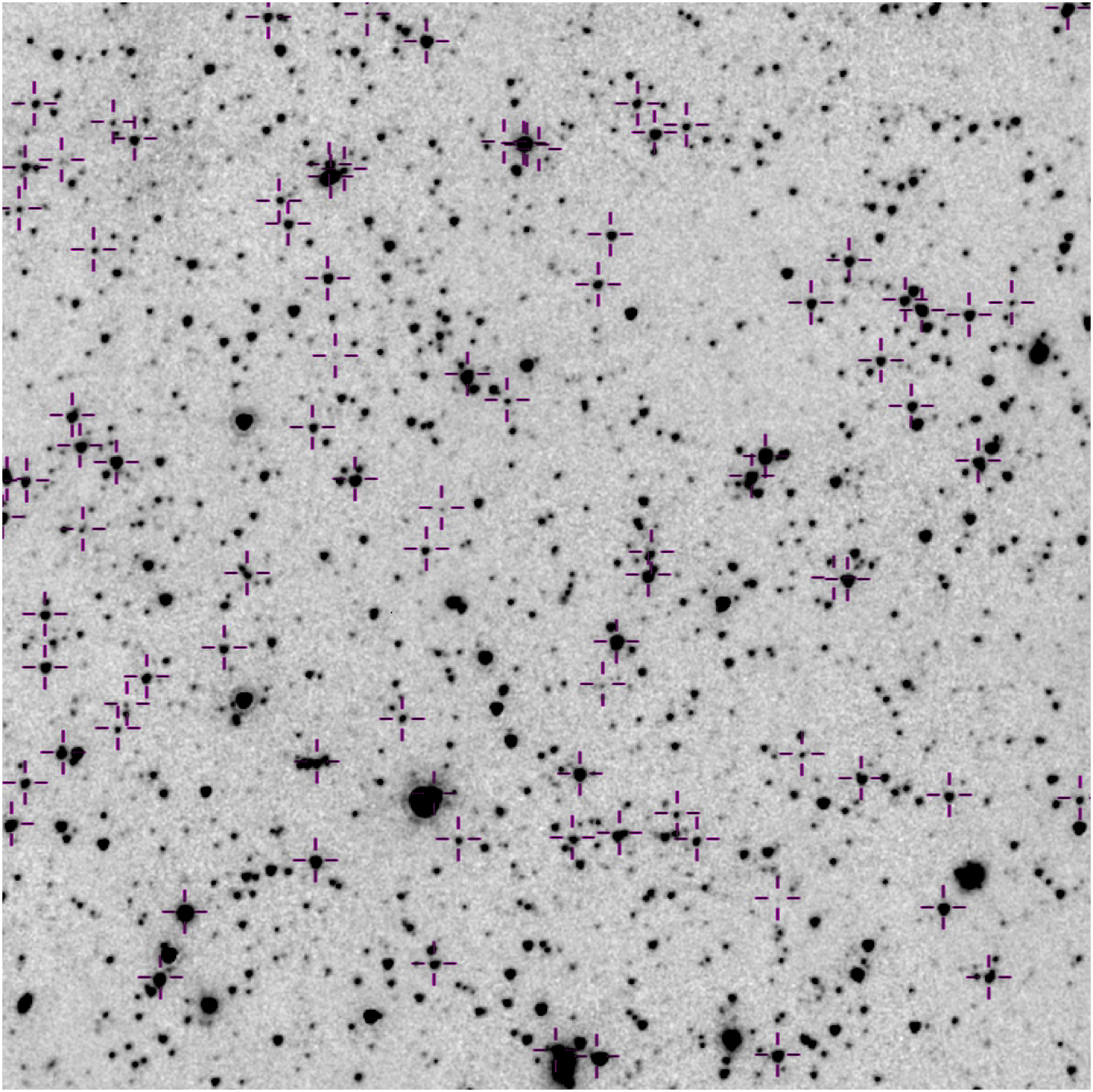}
}
\caption{ 
Radio positions marked by crosses on negative gray scale of the $\lambda = 4.5 \, \mu$m
{\it Spitzer} image.
The area shown is $7\, \farcm 2 \times 7\, \farcm 0$, with north up
and east to the left; this is only a portion of the field imaged.
} 
\label{Radio4.5umFig}
\end{figure*}

Most of our cataloged radio sources are powered by star-forming
galaxies and AGNs that should be visible in sensitive IR images.  Deep
$\lambda = 3.6\,\mu\mathrm{m}$ and $\lambda = 4.5\,\mu\mathrm{m}$ images from
the Spitzer Extragalactic Representative Volume Survey (SERVS)
\citep{Mauduit2012} cover the entire area we imaged at S--band,
although part of the overlapping IR image is blinded by scattered
light from the very bright star GX UMa at J2000 $\alpha =
10^\mathrm{h} \, 46^\mathrm{m} \, 07\,\fs70$,
$\delta = +59^\circ \, 03\arcmin \, 39\,\farcs2$.
Furthermore, the \citet{Mauduit2012} IR
catalog excludes small regions around moderately bright foreground stars in
which galaxies are still visible.  We excluded from our radio/IR
comparisons only the 48 sources in regions that are actually blinded
by bright stars and kept as identification candidates all visible IR
galaxies that had been excluded from the \citet{Mauduit2012} catalog.  At
both $\lambda = 3.6\,\mu\mathrm{m}$ and $\lambda = 4.5\,\mu\mathrm{m}$ the
\citet{Mauduit2012} catalog $5 \sigma$ point-source detection limit is
$S\approx 2\,\mu$Jy and the IR images have FWHM resolution $\theta
\approx 2''$.  The {\it Spitzer} $\lambda = 4.5\,\mu$m image in
Figure~\ref{Radio4.5umFig} shows that nearly all of our radio
source positions (crosses) have  IR identifications.

\begin{figure*}
  \centerline{
  \includegraphics[angle=-0,width=3.6in, trim = -20 140 70 150]{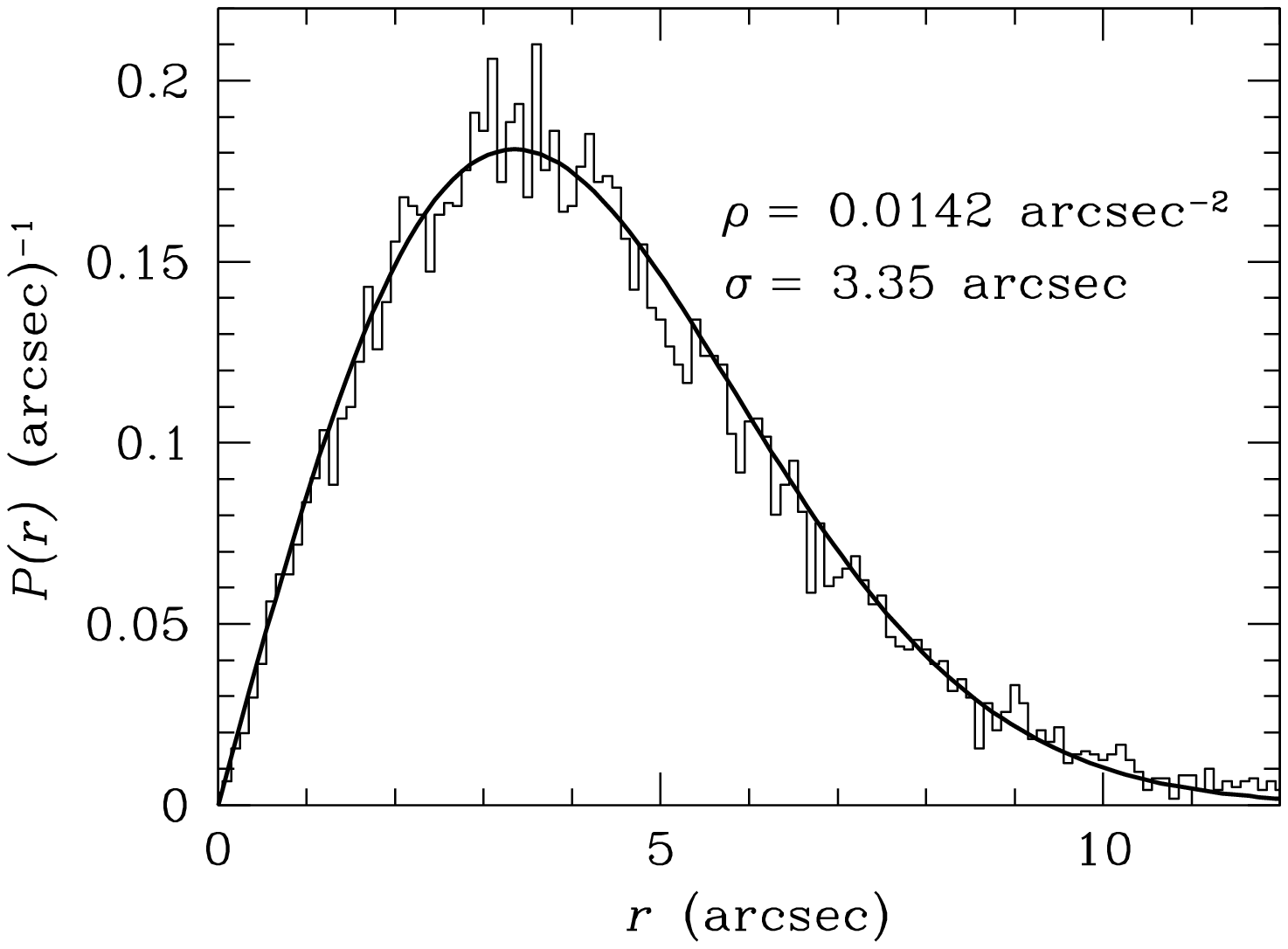}
  \includegraphics[angle=-0,width=3.6in, trim = -20 140 70 150]{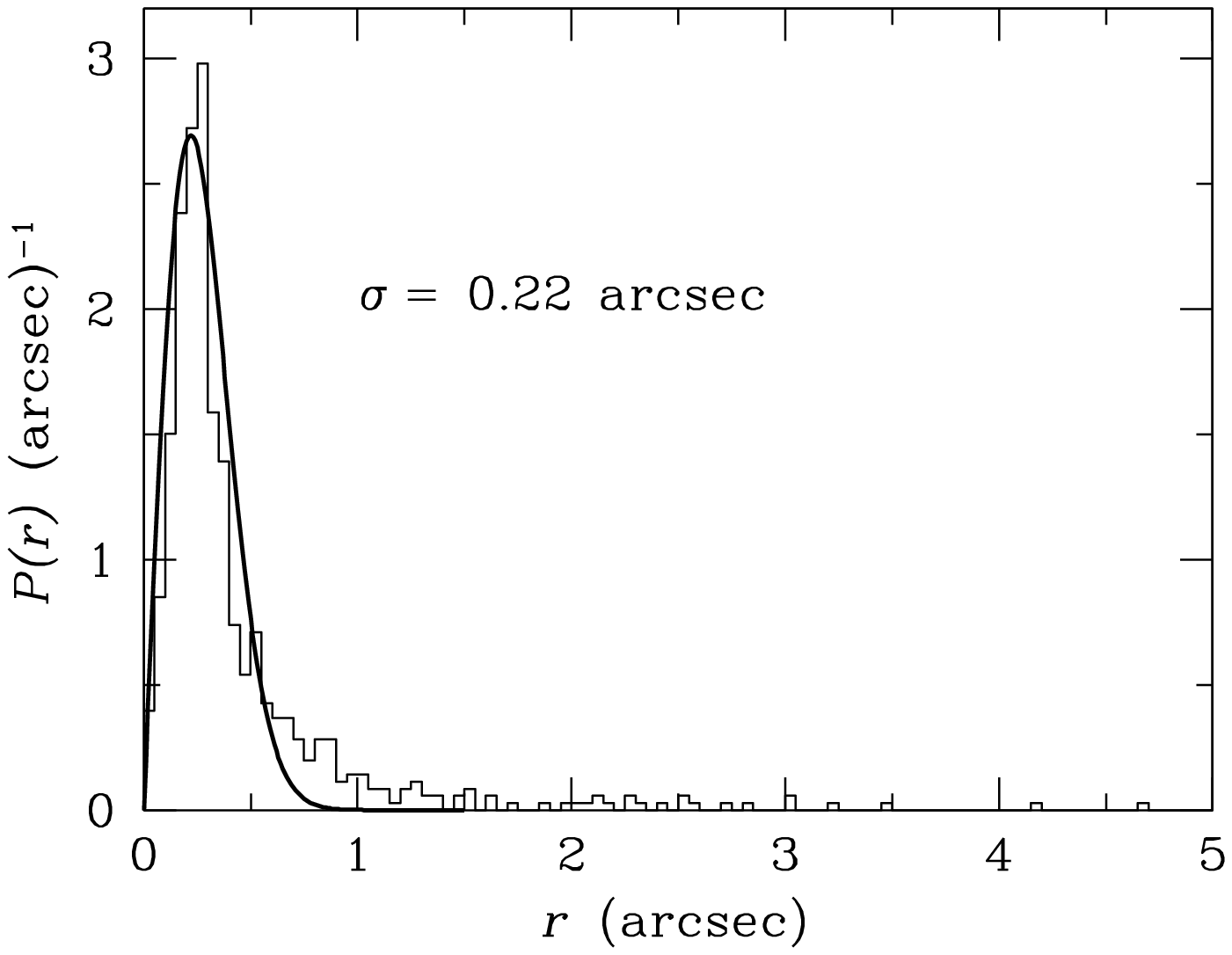}
}
\caption{ Left panel: Histogram of measured radial offsets $r$ between
  points on an arbitrary grid of $120 \times 120$ positions separated
  by $10''$ and their nearest IR neighbors.  The data are well fit by
  the Rayleigh distribution (smooth curve) with rms $\sigma =
  3\,\farcs35$, expected for randomly scattered IR sources with mean sky
  density $\rho = 0.0142 \mathrm{~arcsec}^{-2}$.  Right panel:
  Histogram of measured radial separations $r$ between radio sources
  in Table~\ref{ArcadeACatalog} (excluding the IR-{\rev blinded} 
  fields and the faint sources kept only because they have IR
  counterparts within $1\farcs5$) and their nearest IR neighbors.  The
  core of this distribution can be approximated by a Rayleigh
  distribution with rms $\sigma \approx 0\,\farcs22$, but there is a
tail of somewhat larger offsets, many of which are genuine radio-IR
matches.  Abscissae: Angular distance (arcsec) between a radio
source and its nearest IR neighbor.  Ordinates: Probability density
(arcsec$^{-1}$).}
\label{FalseMatchFig}
\end{figure*}

We identified {\it Spitzer} IR sources with radio sources in
Table~\ref{ArcadeACatalog} on the basis of position coincidence: the
IR source nearest to the radio source was accepted as the
identification if it lies within our maximum search radius
$r_\mathrm{s}$. The probability that an unrelated IR source will
incorrectly be identified depends on the radio and IR position
uncertainties and on the sky density $\rho$ of IR identification
candidates.  The left panel of Figure~\ref{FalseMatchFig} shows a
histogram of radial distances $r$ to the IR sources nearest to an
arbitrary grid of $120 \times 120 = 14400$ positions spaced by $10''$
in right ascension and declination. The histogram is  well approximated
by the expected Rayleigh distribution
\begin{equation}
  P(r) = \frac {r}{\sigma^2} \exp\biggl(- \frac{x^2}{2\sigma^2}\biggr)
  = 2 \pi \rho r \exp(-\pi \rho r^2)~,
\end{equation}
where $\sigma \approx 3\,\farcs35$ is the fitted rms width of the
distribution and $\rho = (2 \pi \sigma^2)^{-1} \approx 0.0142
\mathrm{~arcsec}^{-2}$ is the implied sky density of IR sources.

This result can be used to calculate how strongly having
IR companions within $1\farcs5$  confirms the reality
of the 40 faint ($4 \leq \mathrm{SNR} < 5$) sources found only on the $\theta = 3''$
radio image. The cumulative Rayleigh
distribution 
\begin{equation}\label{eqn:RayleighPcum}
  P(<r) = 1 - \exp\biggl( - \frac{r^2}{2 \sigma^2}\biggr)~
\end{equation}
specifies the probability that an unrelated IR source lies within
a distance $r$ from any point on the sky.  For $\sigma = 3\,\farcs35$,
the probability that a spurious radio source would have an IR companion within
$1\farcs5$ is $P(< 1\farcs5) \approx 0.095$, so an IR confirmation boosts
the reliability of a $4 \leq \mathrm{SNR} < 5$ radio source by a factor of
$0.095^{-1} \sim 10$.

The offsets $r$ of most genuine radio/IR identifications should have
a roughly Rayleigh distribution whose rms is the
quadratic sum of the radio position error, the IR position error, and
any radio-IR offset intrinsic to the host galaxy.  The
distribution of IR offsets from the radio positions of the
752 radio sources \emph{not} confirmed only by an IR source lying
within $1\farcs5$ is shown by the
histogram in the right panel of Figure~\ref{FalseMatchFig}, and the
continuous curve fitting most sources is a Rayleigh distribution with
$\sigma \approx 0\,\farcs22$.  However, there is a tail of sources
with offsets too large to be consistent with this Rayleigh distribution
yet too small to be explained by the Rayleigh distribution of unrelated sources
shown in the left panel.  Such tails are not rare
\citep[e.g.,][]{Murphy2017} and can be attributed to a few sources
with larger combined position errors, extended galaxies with
genuine IR-radio offsets, {\rev sources in clusters} and a small
contamination by unrelated IR sources.

To determine the optimum search radius $r_\mathrm{s}$ that will accept
most genuine identifications and minimize contamination by unrelated IR
sources, we exploited the fact that all unrelated sources should
obey the Rayleigh offset distribution with $\sigma =
3\,\farcs35$ shown in the left panel of Figure~\ref{FalseMatchFig}.
The fraction of background sources with $r > \sigma = 3\,\farcs35$ is
$P(>r) = 1 - P(<r) \approx 0.606$.  On the conservative assumption
that all 10 sources with $r > \sigma = 3\,\farcs 35$ in our catalog
are unrelated to their {\rev IR} neighbors, the total number of unrelated
IR sources with \emph{any} $r > 0$ should be $N_\mathrm{u} \approx (10 \pm 3)
/ 0.606 \approx 16.5 \pm 5$.  Excluding the 40 faint sources cataloged only
because they have IR identifications and the 48 unidentifiable sources
in regions overwhelmed by bright IR stars, Table~\ref{ArcadeACatalog}
contains an IR-unbiased sample of 704 radio sources of which
$\approx (704 - 16.5) / 704 = 97.7 \pm 0.8$\% have true IR identifications
stronger than $S \approx 2 \,\mu$Jy at $\lambda = 4.5 \,\mu$m.  This
high radio/IR identification rate also indicates that $\lesssim 2$\%
of the cataloged radio sources can be spurious.

The expected numbers of background sources in different ranges of $r$
can be calculated from Equation~\ref{eqn:RayleighPcum} and compared
with the observed numbers plotted in the right panel of
Figure~\ref{FalseMatchFig} to estimate the reliability $R$ of an
identification as a function of $r$, as shown in
Table~\ref{table:IDreliability}.  For $r < 1''$, $R \approx 1$.  In
the range $1\arcsec < r < 1\farcs5$, the average number of unrelated
background sources $\langle N_\mathrm{u} \rangle \approx 0.85$ is much
smaller than the observed number $N_\mathrm{o} = 23$ of radio/IR
matches, suggesting that $R \approx (N_\mathrm{o} - \langle
N_\mathrm{u} \rangle) / N_\mathrm{o}) \approx 0.96$.  Thus a search radius
$r_\mathrm{s} = 1\farcs5$ should yield highly reliable radio/IR
identifications.
Table~\ref{table:IDreliability} shows how reliability decreases for
larger separations. {\rev Nevertheless, more} than half of the
``probable'' identifications with $1\,\farcs5 < r < 3\,\farcs5$ appear
to be correct. 

\begin{table}
\caption{IR identification reliability. $R$}
\vskip 0.1in
\begin{center}
\begin{tabular}{cccc}
\hline
\hline
  \colhead{ $~r('')$} & \colhead{$\langle N_\mathrm{u} \rangle$} &
  \colhead{$N_\mathrm{o}$}& \colhead{$R$} \\
\hline
1.0--1.5 & 0.85 & 23 & 0.96 \\
1.5--2.0 & 1.12 & \hphantom{2}8  & 0.86 \\
2.0--2.5 & 1.32 & \hphantom{2}9  & 0.85 \\
2.5--3.0 & 1.44 & \hphantom{2}5  & 0.71 \\
3.0--3.5 & 1.49 & \hphantom{2}4  & 0.63 \\
\hline
\end{tabular}
\end{center}
\break
Notes: For each radio-IR offset range $r$, $\langle N_\mathrm{u} \rangle$
is the average number of unrelated IR sources, $N_\mathrm{o}$
is the observed number of IR sources, and $R$ 
is the estimated identification reliability.
\label{table:IDreliability}
\end{table}

Let $m \equiv r_\mathrm{s} / \sigma$ be the search radius in units of
the rms position error $\sigma \approx 0\,\farcs22$ and define $k
\equiv 1 + 2 \pi \rho \sigma^2 \approx 1.0043$.  Then the completeness
$C$ of our position-coincidence identifications is
\begin{equation}
  C = \frac{ 1 - \exp(-m^2k / 2)}{k} \approx 0.996
\end{equation}
and the reliability $R$ is
\begin{eqnarray}
  R = C \biggl[ \frac{1}{f} - \biggl(1 - \frac{1}{f}\biggr) \exp [m^2(1-k) /2)] - \nonumber \\
    \exp\biggl(-\frac{m^2 k}{2}\biggr) \biggr]^{-1} \approx 0.957 ~~~~~
\end{eqnarray}
\citep{condonetal75}.

\section{Source Size Distribution}
Most of the individual fitted Gaussian sizes produced in the
source-finding process are sufficiently uncertain to yield only upper
limits to the individual deconvolved source sizes.  However, the mean
ratio $\langle S_\mathrm{p}^\mathrm{H} /
S_\mathrm{p}^\mathrm{L}\rangle$ of the fitted peak {\rev brightness}
$S_\mathrm{p}^\mathrm{H}$ {\rev (on the $\theta_\mathrm{H} = 0\,\farcs66$
high-resolution image)} to $S_\mathrm{p}^\mathrm{L}$ {\rev (on the
$\theta_\mathrm{L} = 3\arcsec$ low-resolution image)} depends on the
source solid angle.  If a circular Gaussian source of FWHM diameter
$\phi$ and flux density $S$ is imaged with a beam of FWHM diameter
$\theta$, the source appears on the image as a circular Gaussian with
FWHM diameter $(\theta^2 + \phi^2)^{1/2}$ and peak {\rev brightness}
\begin{equation}
  S_\mathrm{p} = S \biggl( \frac {\theta^2}{\theta^2 + \phi^2} \biggr)~.
\end{equation}
If the circular Gaussian source is imaged with two different resolutions
$\theta_\mathrm{H}$ and $\theta_\mathrm{L}$, the ratio of the image
peak {\rev brightnesses} is 
\begin{equation}\label{eqn:spratio}
  \frac {S_\mathrm{p}^\mathrm{H}}{S_\mathrm{p}^\mathrm{L}} =
  \biggl( 1 + \frac{\phi^2}{\theta_\mathrm{L}^2} \biggr)
  \biggl( 1 + \frac{\phi^2}{\theta_\mathrm{H}^2} \biggr)^{-1}~.
\end{equation}
Appendix C of \citet{Murphy2017} gives the $S_\mathrm{p}^\mathrm{H} / S_\mathrm{p}^\mathrm{L}$
ratio for a source with the exponential brightness profile typical of spiral galaxies.
Equation~\ref{eqn:spratio} can be solved for the source size $\phi$:
\begin{equation}\label{eqn:sourcesize}
\phi= \Biggl[ \frac {\theta_\mathrm{L}^2 \theta_\mathrm{H}^2
( S_\mathrm{p}^\mathrm{L} -S_\mathrm{p}^\mathrm{H} )}
{\theta_\mathrm{L}^2 S_\mathrm{p}^\mathrm{H} - \theta_\mathrm{H}^2 S_\mathrm{p}^\mathrm{L}}
\Biggr]^{1/2}~.
\end{equation}
Even if a source does not have a circular Gaussian brightness distribution,
Equation~\ref{eqn:sourcesize} defines what we call its equivalent
source circular Gaussian FWHM.

\begin{figure*}
\centerline{
\includegraphics[width=4.0in,angle=-90]{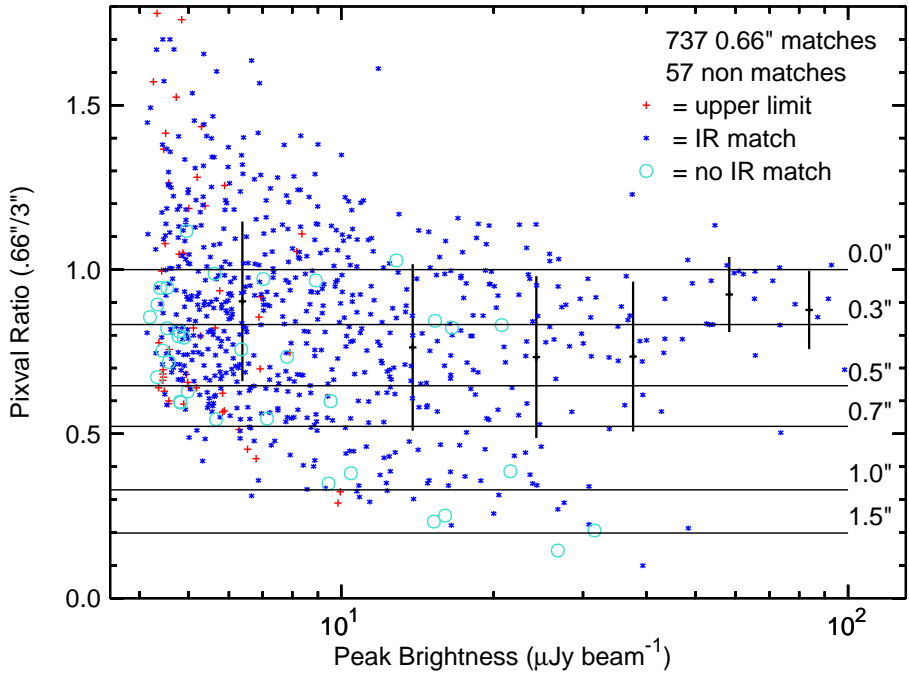}
}
\caption{
Distribution of $S_\mathrm{p}^\mathrm{H} / S_\mathrm{p}^\mathrm{L}$
as a function of $S_\mathrm{p}^\mathrm{L}$ peak {\rev brightness}.
Blue asterisks represent sources with IR matches within $1\,\farcs5$, cyan
circles are sources
with no IR match and small red pluses are  $3\sigma$ upper limits for
sources not detected at $0\,\farcs66$ resolution.
Extended vertical bars are {\rev peak brightness}-bin averaged values with
the vertical extent showing the  rms scatter $\sigma_\mathrm{pop}$ within the population.
Horizontal solid lines mark the expected ratio of circular Gaussians
with FWHM of the labeled size.
{\rev Outliers are excluded from the bin statistics.}
The large number of points with a ratio in excess of 1.0 at low {\rev
peak brightness} is consistent with the noise.
}
\label{RatioVFlux}
\end{figure*}

Equations~\ref{eqn:spratio} and \ref{eqn:sourcesize}
allow us to estimate the statistical properties of the source
sizes in our sample.  The distribution of
$S_\mathrm{p}^\mathrm{H} / S_\mathrm{p}^\mathrm{L}$ as a function of
$S_\mathrm{p}^\mathrm{L}$ is given in Figure~\ref{RatioVFlux}. Sources with and
without IR counterparts are shown by different symbols. Horizontal
lines give the expected values for circular Gaussians of various FWHM
diameters $\phi$.

The cataloged sources were separated into {\rev peak brightness} bins in which
the bin-averaged {\rev peak brightnesses} and peak {\rev brightness}
ratios are plotted in Figure \ref{RatioVFlux} with ``error bars''
giving the rms ratio for the population in each bin. 
The {\rev peak brightness} bin statistics are given in Table~\ref{SizeTab}.
{\rev Outliers further from from the initial mean by more than 2
$\sigma$ were excluded from the analysis of bin popluations.}


At {\rev peak brightnesses} below the 20--30~$\mu$Jy {\rev beam$^{-1}$}  bin,
the distribution of {\rev peak brightness} ratios appears truncated on
the low end by limited sensitivity. 
The 20--30~$\mu$Jy {\rev beam$^{-1}$}  and higher {\rev peak
brightness} bins in Table~\ref{SizeTab} consistently give an
equivalent source circular Gaussian FWHM {\rev of}  $\theta \approx 0\,\farcs3$. 
\begin{table}
\caption{Measured average ratios of peak flux densities.}
\vskip 0.1in
\begin{center}
\begin{tabular}{rccccc} 
\hline
\hline
No.  & $\langle S_\mathrm{p}^\mathrm{L} \rangle$ & $\langle
 S_\mathrm{p}^\mathrm{H}/S_\mathrm{p}^\mathrm{L} \rangle$ &
 $\sigma_{\rm pop}$  & FWHM  & err\\
           &  $\mu$Jy    &           &                & $\arcsec$ & $\arcsec$ \\
\hline
\rev
428 & \hphantom{2} 6.4 & 0.76 & 0.18 & 0.39 &0.19 \\
165 & 13.9 & 0.71 & 0.19 & 0.43 & 0.21 \\
52  & 24.3 & 0.69 & 0.25 & 0.45 & 0.27 \\
31  & 37.4 & 0.77 & 0.18 & 0.38 & 0.19 \\
14  & 58.3 & 0.87 & 0.11 & 0.26 & 0.13 \\
9   & 83.8 & 0.93 & 0.07 & 0.20 & 0.09 \\
\hline
\end{tabular}
\end{center}
\hfill\break
Notes:
``No.'' is the number of sources in the bin, {\rev excluding outliers}, 
$\langle S_\mathrm{p}^\mathrm{L} \rangle$ is  
the average peak {\rev brightness} at $3\arcsec$ resolution, $\langle S_\mathrm{p}^\mathrm{H} /
S_\mathrm{p}^\mathrm{L} \rangle$ is the average peak {\rev brightness} ratio,
$\sigma_{\rm pop}$ is the rms ratio of the bin sample, ``FWHM'' is the
equivalent half-power diameter of a circular Gaussian with the ratio of the bin
average, {\rev and ``err'' is the estimated 1 $\sigma$ error of ``FWHM''.}
\label{SizeTab}
\end{table}

\begin{figure*}
\centerline{
\includegraphics[width=4.0in,angle=-90]{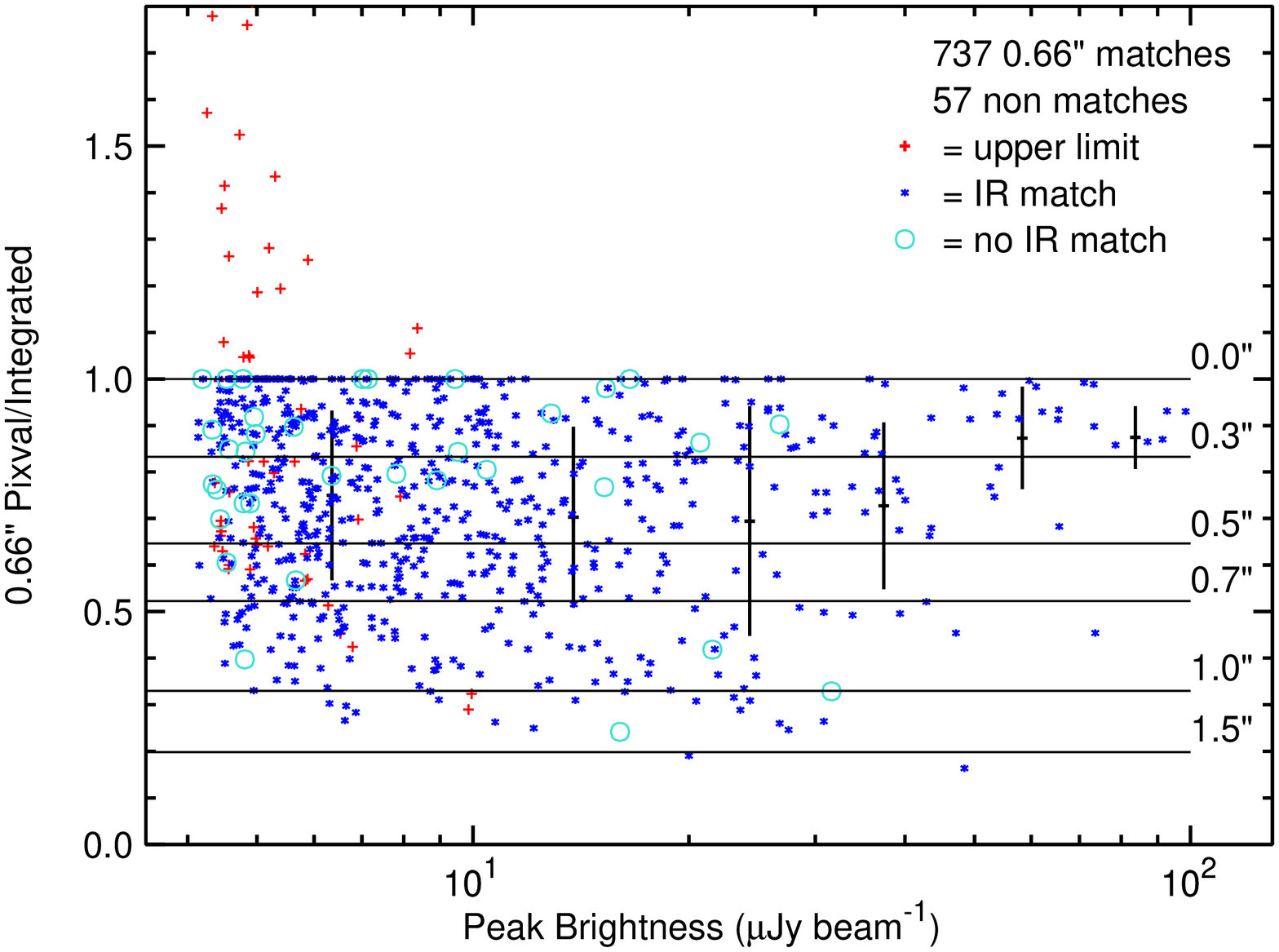}
}
\caption{ Like Figure~\ref{RatioVFlux}, but for the ratio
  $S_\mathrm{p}^\mathrm{H} / S^\mathrm{H}$ versus  peak  {\rev brightness}
  $S_\mathrm{p}^\mathrm{H}$ in the
  $0\,\farcs66$ resolution image.  Red plus symbols are for sources
  not detected at $0\,\farcs66$ resolution, where the ratio is the
  $3\sigma$ upper limit to the $3\arcsec$ {\rev peak brightness} and the flux
  density is {\rev the peak brightness} at $3\arcsec$ resolution.  The
  plotted ratio is equivalent to the ratio of the fitted beam area to
  the restoring beam area; fitting constraints limit this to a ratio
  of 1.0.  
  The lowest {\rev peak brightness} bin is biased low with respect to Figure
  \ref{RatioVFlux} but the higher {\rev peak brightness} bin averages are
  comparable in spite of being derived solely from the $0\,\farcs66$
  data.  }
\label{RatioAOnly}
\end{figure*}

Figure~\ref{RatioVFlux} shows that the detected sources are generally
marginally resolved at $0\,\farcs66$ resolution.
A more sensitive (although biased) estimate of the source size is the
ratio of the peak {\rev brightness} to the estimated integrated flux
density from only the $0\,\farcs66$ data.
The integrated flux density is the Gaussian peak times the ratio of
the fitted beam area to the CLEAN restoring beam area. 
Because the Gaussian fits were constrained to give a fitted beam no
smaller than the CLEAN restoring beam, this estimate will be biased for
point, or nearly unresolved sources.
The plot of peak {\rev brightness} to integrated flux densities is
shown in Figure \ref{RatioAOnly}.

\section{Completeness}
\begin{figure*}\centerline{
\includegraphics[width=4.0in,angle=-90]{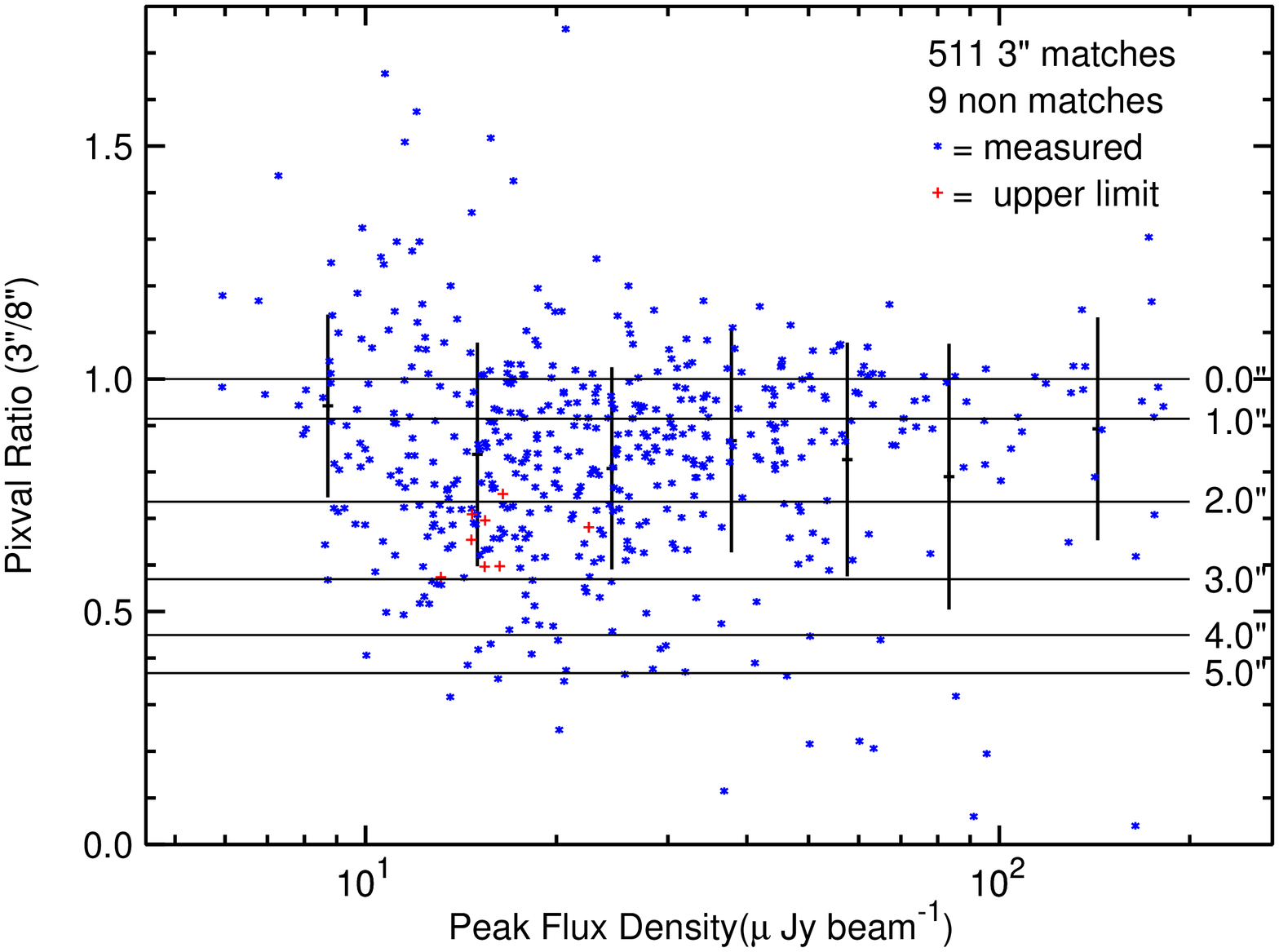}
}
\caption{
Like Figure~\ref{RatioVFlux}, but for the ratio of the peak {\rev brightness} at
3$\arcsec$ resolution to that at 8$\arcsec$ resolution versus the
8$\arcsec$ peak {\rev brightness}.
All but nine of the  8$\arcsec$ resolution sample were detected, but the
peak {\rev brightness} are systematically below 1.0.
}
\label{LowResRatio}
\end{figure*}

The completeness of the sample of sources selected at $3\arcsec$ resolution
given  in Table~\ref{ArcadeACatalog} was evaluated using a
comparison with the 8$\arcsec$ resolution image of \cite{Paper1}. 
A catalog of sources generated from the 8$\arcsec$ resolution image was
cross--matched with the catalog generated from the 3$\arcsec$ resolution image.
Out of the 503 8$\arcsec$ resolution entries stronger than $5\sigma$,
eight did not have matches at 3$\arcsec$ resolution.
Of these, four were  extended lobes of nearby bright sources,
plausibly undetected at 3$\arcsec$ resolution, and one was in the
{\rev masked} region of the 3$\arcsec$ image.
This suggests that the sample given in Table~\ref{ArcadeACatalog} is
$\approx 98.5$\% complete.
The ratios of the peak flux densities at 3$\arcsec$ to 8$\arcsec$ resolution for this
sample are given in Figure~\ref{LowResRatio}.
The analysis displayed in Figure~\ref{LowResRatio} suggests a larger ``typical''
source size ($\sim 1\,\farcs0$) than Figure~\ref{RatioVFlux}, but this is a
relatively small fraction of the 3$\arcsec$ resolution.
{\rev The confusion ``noise'' in the 8$\arcsec$ image will also bias
the flux density ratio lower, especially at the low flux densities.}

Of the 792 entries in Table~\ref{ArcadeACatalog}, 732 were detected in
the $0\,\farcs66$ image {\rev at SNR$>$3} while 60 were not.
Thus, the high-resolution images have a 92\% detection rate of the $3\arcsec$
resolution sources, or approximately 91\% of sources detectable at $8\arcsec$
resolution. 
The sources that were not detected at $0\,\farcs66$ resolution are among
the weakest in the sample.

\begin{figure}\centerline{
\includegraphics[width=2.75in,angle=-90]{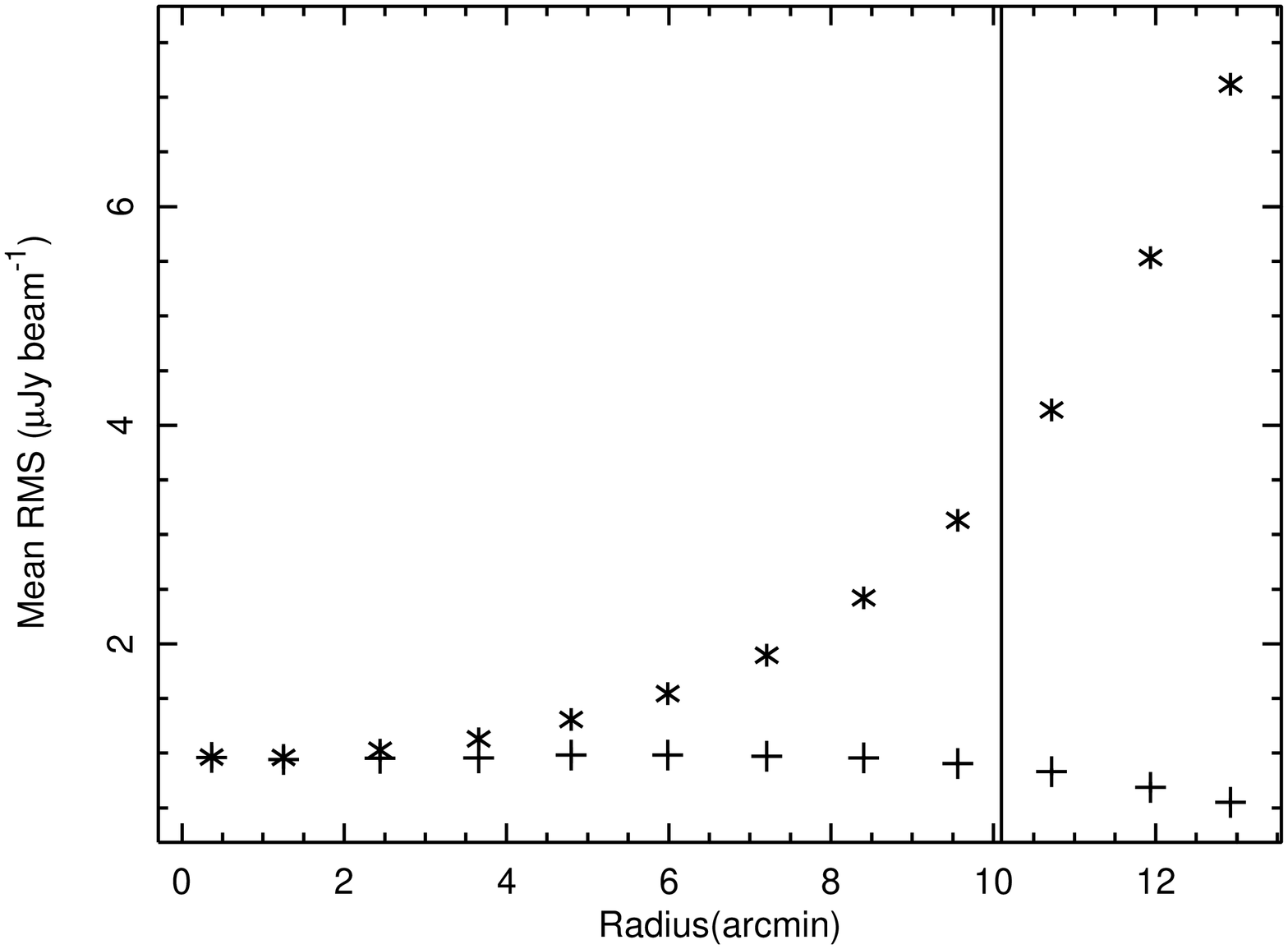}
}
\caption{
\rev
Average source {\rev rms} in the $0\,\farcs66$ image as a function of
distance from the pointing center.
Plus symbols indicate the values without antenna gain correction, while
asterisks are after correction.
The vertical line indicates the maximum distance from which sources
were included in Table~\ref{ArcadeACatalog}.
}
\label{RMS}
\end{figure}

{\rev The completeness as a function of the flux density is a function
of the local image {\rev rms} and thus a function of the antenna gain.
Since only a single pointing was used, the noise is only (nearly)
stationary in the image prior to correction for the antenna gain.
The noise quoted in Table~\ref{ArcadeACatalog} was measured in a
$201\times201$ pixel box around {\rev each} source, scaled by the inverse of
the antenna gain. 
The average {\rev rms} with and without antenna gain corrections as a
function of distance from the pointing center for the  $0\,\farcs66$
image is shown in Figure~\ref{RMS}.
}

\section{Simulations\label{simulations}}
\begin{figure*}
\centerline{
\includegraphics[width=4.0in,angle=-90]{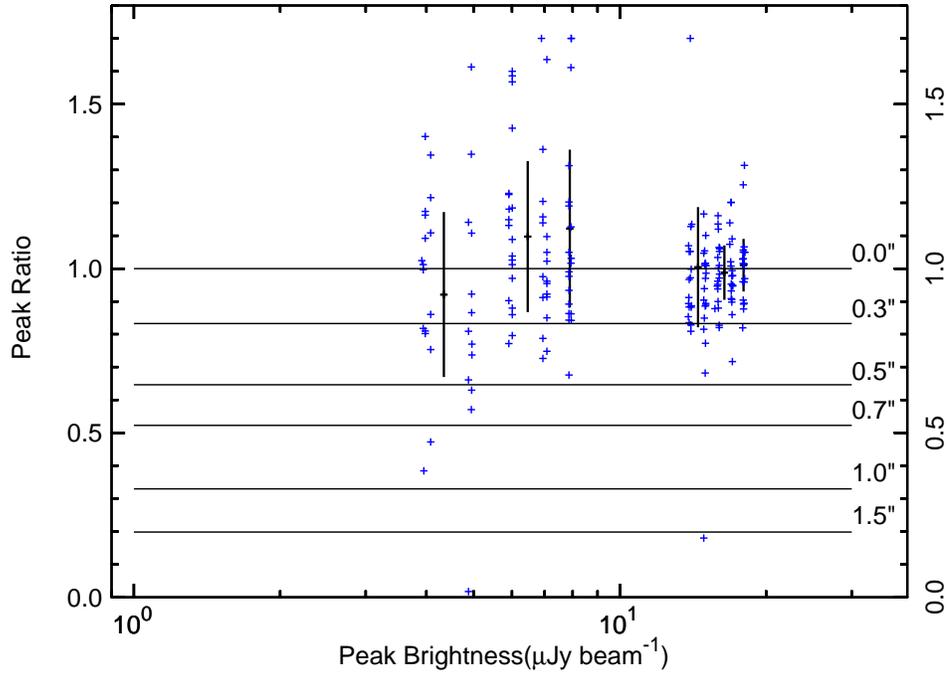}
}
\centerline{
\includegraphics[width=4.0in,angle=-90]{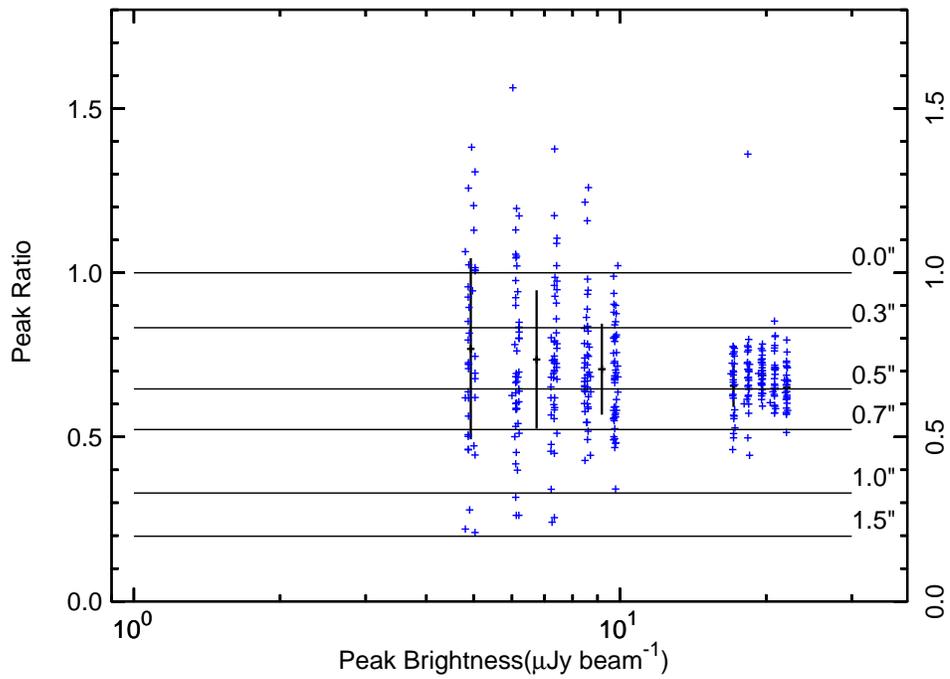}
}
\caption{
Like Figure~\ref{RatioVFlux}, but for artificial sources
added to the data and imaged.
The abscissa is the model $S_\mathrm{p}^\mathrm{L}$.
{\rev The upper} plot has point sources, {while the lower has
}circular Gaussians of $0\,\farcs5$ FWHM. 
Bin averages and rms deviations of the population are shown by the
black vertical bars. 
}
\label{FakeVFlux}
\end{figure*}

In order to estimate the effects of imaging and image analysis on the
distribution of the apparent sizes of sources, simulated point 
and $0\,\farcs5$ circular Gaussian sources with a range of flux
densities were added to the visibility data sets.
These were then imaged and the response in the derived image measured.
In each of several simulations 144 artificial sources of a
given size were added in a hexagonal grid;
the location of the grid did not take into account the locations of
real sources. 
The distributions of point and $0\,\farcs5$ FWHM Gaussians are shown in Figure
\ref{FakeVFlux}.   

The simulated measurements tend to scatter around the expected ratio,
although with less dispersion than seen in the actual source
population in Figure~\ref{RatioVFlux}.
Outliers at the higher flux densities appear when the artificial source
is located near a real source and the two are blended at 3$\arcsec$ resolution.
Bin averages and rms values of the populations are shown in Figure
\ref{FakeVFlux} and are given in Table~\ref{FakeTab}.
{\rev Outliers further from the initial mean by more than 2
$\sigma$ were excluded from the analysis of bin popluations.}
Table~\ref{FakeTab} suggests that our analysis may statistically
underestimate the sizes of the  faintest sources.
This effect is {\rev possibly} the result of the bias in the Gaussian fitting
of the high-resolution image towards the peak in the source plus noise,
which will bias the derived peak {\rev brightness} high.
\begin{table}
\caption{Simulation flux bin ratio averages.}
\vskip 0.1in
\begin{center}
\begin{tabular}{rccccrr} 
\hline
\hline
No.  & $\phi$ & $\langle S_\mathrm{p}^\mathrm{L} \rangle$ &
$\langle S_\mathrm{p}^\mathrm{H} / S_\mathrm{p}^\mathrm{L} \rangle$ & $\sigma_\mathrm{pop}$  & FWHM & err\\
    &  $\arcsec$     &  $\mu$Jy    &           &                & $\arcsec$ & $\arcsec$ \\
\hline
\rev
27 & 0.0 &  4.4  & 0.91 & 0.25 & 0.21 &  0.33 \\
36 & 0.0 &  6.5  & 1.07 & 0.23 & 0.00 & 24.6  \\
16 & 0.0 &  7.9  & 1.02 & 0.24 & 0.00 & 25.7  \\
36 & 0.0 & 14.5  & 0.93 & 0.18 & 0.19 &  0.26 \\
36 & 0.0 & 16.4  & 0.99 & 0.08 & 0.08 &  0.26 \\
17 & 0.0 & 18.0  & 0.98 & 0.08 & 0.11 &  0.18 \\
36 & 0.5 &  4.9  & 0.74 & 0.28 & 0.40 &  0.29 \\
88 & 0.5 &  6.7  & 0.72 & 0.21 & 0.43 &  0.23 \\
89 & 0.5 &  9.2  & 0.69 & 0.14 & 0.46 &  0.15 \\
25 & 0.5 & 17.1  & 0.66 & 0.09 & 0.49 &  0.10 \\
85 & 0.5 & 19.6  & 0.68 & 0.06 & 0.47 &  0.07 \\
26 & 0.5 & 22.0  & 0.65 & 0.05 & 0.50 &  0.06 \\
\hline
\end{tabular}
\end{center}
\hfill\break
Notes:
``No,'' is the number of simulated sources in the bin, {\rev excluding outliers}. 
The column labeled $\phi$ in Table~\ref{FakeTab} gives the FWHM of the
simulated source, while ``FWHM'' gives the equivalent circular Gaussian
half-power diameter corresponding to the bin average ratio,
{\rev and ``err'' is the estimated 1 $\sigma$ error of ``FWHM''.}
\label{FakeTab}
\end{table}

A comparison between real and simulated sources of the statistics for
the ratio $S_\mathrm{p}^\mathrm{H} / S_\mathrm{p}^\mathrm{L}$ (for
sources strong enough that the distribution is not truncated by
sensitivity) can help constrain the range of source effective sizes.
The simulated source results will include all of the noise and image
processing artifacts suffered by the real sources, except for
calibration and antenna pointing errors.
The bulk of the sources not detected at the higher resolution are at
the lower flux densities and are likely not detected due to
sensitivity.  
The following is a general analysis of the source
population at these flux density levels.

The real sources in the well-sampled range of 10--50~$\mu$Jy
beam$^{-1}$ have a typical population rms $\sigma_\mathrm{pop} = 0.26$
in this ratio (Table~\ref{SizeTab}).  The simulated sources in
Table~\ref{FakeTab} typically have $\sigma_\mathrm{pop} = 0.10$ in
this {\rev peak brightness} range for both point and $0\,\farcs5$ circular
Gaussians.  The expected ratio for a $0\,\farcs5$ Gaussian is
$S_\mathrm{p}^\mathrm{H} / S_\mathrm{p}^\mathrm{L} = 0.65$ and for a
point source is $S_\mathrm{p}^\mathrm{H} / S_\mathrm{p}^\mathrm{L} =
1.0$.
{\rev  The difference between the real and simulated source samples is
that a range of sizes is expected for the real sources, whereas the
simulations used only a single source size.
The difference in the scatter of the peak brightness ratios between
real and simulated sources is consistent with this expectation.
}

Thus the scatter in $S_\mathrm{p}^\mathrm{H} /
S_\mathrm{p}^\mathrm{L}$ is significantly less than the difference
between an unresolved and a $0\,\farcs5$ source.  This means that the
real source size distribution must be relatively tightly clustered about
the typical $\phi \approx 0\,\farcs3$ equivalent circular Gaussian.
Assuming the measured $\sigma_\mathrm{pop}$ values are {\rev the} rms scatter of
a Gaussian distributed population, a simple analysis of the ratio
statistics gives a distribution with a mean value of $\langle \phi
\rangle = 0\,\farcs 3 \pm 0\,\farcs1$ and an rms scatter $\lesssim
0\,\farcs3$ of the distribution.



\section {FIR/radio correlation}
\begin{figure*}
\centerline{
\includegraphics[width=5.0in,angle=-90]{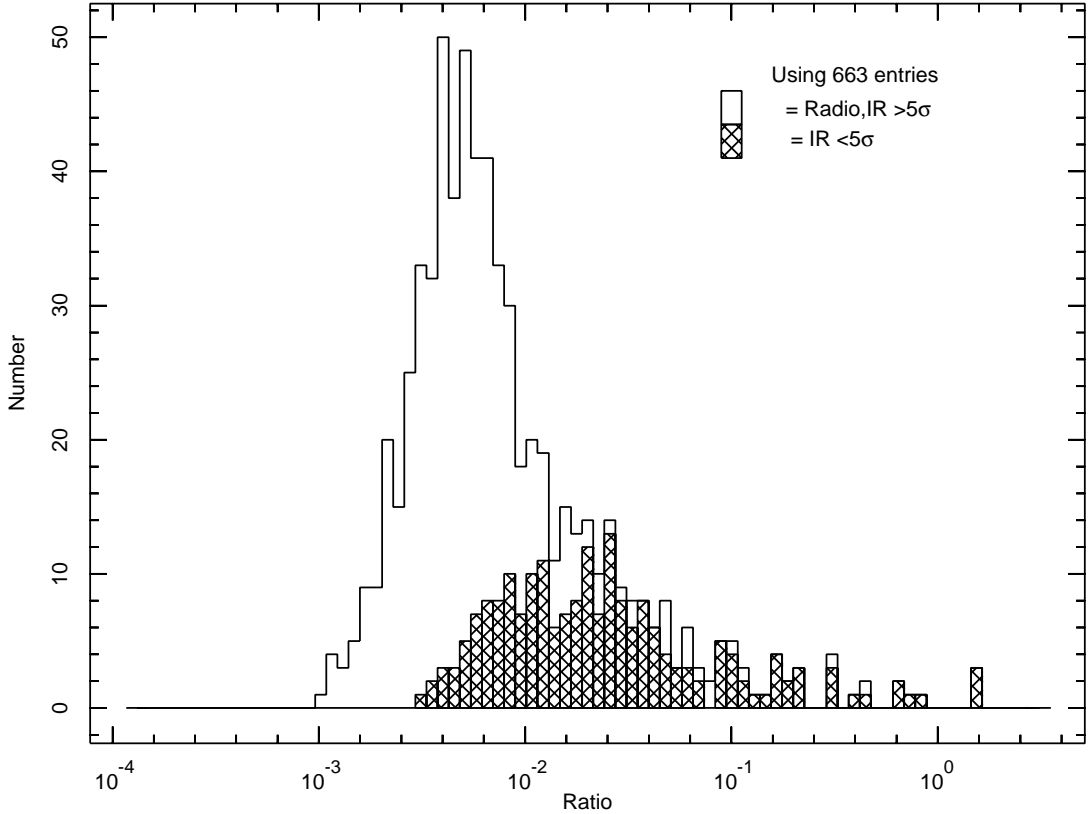}
}
\caption{
Histogram of the ratio of the 3~GHz flux density to the $\lambda
= 160\,\mu$m {\it Herschel} flux density for the sources in Table
\ref{ArcadeACatalog}.
Hatched areas indicate sources without a solid {\it Herschel} detection.
}
\label{ArcadeHerschel160}
\end{figure*}

The positions of radio sources in Table~\ref{ArcadeACatalog} brighter
than $5\sigma$ were examined in deep {\it Herschel} $\lambda =
160\,\mu$m images \citep{Herschel160}.  The {\it Herschel} image is
confusion limited, so only sources brighter than {\rev 5} times the
$0.4\,\mu\mathrm{Jy~pixel}^{-1}$ rms noise were considered real.
Aperture photometry was performed on the $160\,\mu$m image by summing
all pixels within {\rev a radius of three pixels} and multiplying by
1.195.  
This correction factor was derived by a comparison on the brightest
source in the field with a much larger aperture.  The results are shown in
Figure~\ref{ArcadeHerschel160}, with upper limits indicated by the
{\rev hatched} areas.  The bulk of the sources with $160\,\mu$m detections are
within a fairly narrow range of $\log [\langle S(160\,\mu \mathrm{m})
    / S(3\,\mathrm{GHz})] = 2.3$ in the observer's frame, or $\log
  [\langle S(80\,\mu \mathrm{m}) / S(1.5\,\mathrm{GHz})] = 2.3$ in
    the source frame at $z \sim 1$, indicating that most obey the
    FIR/radio correlation typical of star-forming galaxies.

\section {Polarization}
The A-configuration data were imaged in Stokes {\it I, Q}, and
{\it U}.
Sources with polarized emission in excess of twice the polarized rms
after bias correction are listed in Table~\ref{PolnTab}. 
Three have FR I/II morphologies, but another three are either unresolved or
marginally resolved.
Few sources were bright enough that a 1\% polarization could have been
detected.

\begin{table*}
\caption{Polarized sources}
\vskip 0.1in
\begin{center}
\begin{tabular}{cccccl} 
\hline
\hline
 J2000 $\alpha$     & J2000 $\delta$  &  $I$  & $P$  & EVPA & Comment \\
              &            & $\mu\mathrm{Jy~beam}^{-1}$& $\mu\mathrm{Jy~beam}^{-1}$&
               $^\circ$ & \\
\hline
 10 45 25.972 & 58 58 45.06 & 8.3 $\pm$ 0.9 &   3.7 $\pm$  1.7 &   ~+1.8 $\pm$  9.4 & Weak, isolated \\
 10 45 39.783 & 58 57 29.80 & 493 $\pm$ 16 &  14.4 $\pm$ 6.0 &   ~+5.1  $\pm$  4.4 & Part of FR I \\
 10 45 48.960 & 58 54 07.86 & 8.8 $\pm$  1.2 &   5.9  $\pm$  2.4 &  +89.8  $\pm$  8.3 & Extended AGN \\
 10 46 16.140 & 59 04 30.02 & 29.5 $\pm$ 1.2 & 3.8 $\pm$ 1.6 & $-$38.9  $\pm$  8.4 & Isolated, small \\
 10 46 23.967 & 59 06 10.12 & 178 $\pm$ 7\hphantom{0} & 30 $\pm$ 9 & $-29.3$ $\pm$ 2.9 & FR II N Lobe \\
 10 46 24.494 & 59 04 48.44 & 73.7 $\pm$ 4.0 & 9.3 $\pm$ 3.0 & $-$67.4  $\pm$  5.0 & FR II S Lobe \\
 10 46 24.847 & 59 04 45.97 & 3979 $\pm$ 126 & 592 $\pm$ 9 &   ~+0.1  $\pm$  0.1 & FR II S Lobe \\
 10 46 44.540 & 59 01 16.26 & 207 $\pm$ 7 & 12.3 $\pm$ 2.6 & $-$18.1 $\pm$ 3.7 & Isolated, small \\
\hline
\end{tabular}
\end{center}
\hfill\break\
Notes:
$I$ is the {\it Stokes} I peak flux density followed by
its rms error, $P$ is the  polarized peak flux density
{\rev ($\sqrt{Q^2+U^2}$)}
followed by its error,  and {\rev $EVPA$} is the electric field position angle measured
east from north.
\label{PolnTab}
\end{table*}

\section{Extended Sources\label{extended}}
Six sources are sufficiently complex that they are
not adequately described by a superposition of Gaussian components.
They all appear to be AGN-driven FR~I and FR~II sources.  They are
summarized in Table \ref{ExtendedTab}.
\begin{table}
\caption{Extended Sources.}
\vskip 0.1in
\begin{center}
\begin{tabular}{ccrr} 
\hline
\hline
J2000 $\alpha$  & J2000 $\delta$ & ${\rm LAS}$ & $S$~ .\\
           &            & $\arcsec$~~   & $\mu$Jy    \\
\hline
 10 45 37.189 &  59 09 45.48 &  2.8 &   397\\
 10 45 39.783 &  58 57 29.79 & 12.1 &  1877\\
 10 45 49.226 &  58 54 12.59 &  6.7 &   392 \\
 10 46 24.005 &  59 05 22.29 & 86.4 &  8617 \\
 10 46 33.215 &  58 58 15.42 &  2.4 &   239 \\
 10 46 38.337 &  58 54 20.31 & 17.4 &   575 \\
\hline
\end{tabular}
\end{center}
\hfill\break
Notes:
${\rm LAS}$ is the largest angular size, $S$ is the
integrated flux density.
\label{ExtendedTab}
\end{table}

\section{\rev Discussion}
The median size {\rev derived for the faint source population} is
consistent with that expected for a  
population dominated by star-forming galaxies.  The ``effective
radius'' $r_\mathrm{e}$ of a galaxy is the radius enclosing half the
emitted radiation.  For an exponential disk, $r_\mathrm{e} \approx
\phi / 2.43$ \citep{Murphy2017}, so our star-forming galaxies have
$\langle r_\mathrm{e} \rangle \approx 0\,\farcs 12 \approx 1
\mathrm{~kpc}$ for $z \sim 1$.  This is somewhat larger than the
$\langle r_\mathrm{e} \rangle = 0\,\farcs 069 \pm 0\,\farcs 013$
reported by \citep{Murphy2017}, whose $\nu = 10 \mathrm{~GHz}$ sources
have a larger thermal fraction and are less broadened by cosmic-ray
diffusion, and which were selected from an image with higher
resolution ($\theta = 0\,\farcs22$).  
{\rev On the other hand, our} effective radius agrees well
with the stacked H$\alpha$ emission-line effective radius for
high-mass galaxies at $z \sim 1.4$, after it was corrected for dust
extinction \citep{nel16}.

The sources in this study have a slightly smaller median angular size than
galaxies selected at submm wavelengths, for sizes measured
with ALMA observations in the far-infrared or with the VLA at
3~GHz. \citet{Hodge16} estimated a mean effective radius $\langle r_\mathrm{e}\rangle =
1.8\pm 0.2$~kpc of dust emission at 345~GHz in 16 submm galaxies at
$\langle z \rangle \sim 2.5$, and \citet{Simpson15} found
$\langle r_\mathrm{e} \rangle = 1.2 \pm 0.1$~kpc for  52 submm-selected galaxies, also at
345~GHz.  \cite{Miettinen17} reported a  mean Gaussian FWHM size 
$4.6\pm 0.4$~kpc (corresponding to $\langle r_\mathrm{e}
\rangle \approx 1.9$~kpc) at 3~GHz for a sample of 115 submm-selected
galaxies detected in the COSMOS VLA radio survey \citep{Smolcic17}.
However, all three of these studies are dominated by submm-selected
galaxies with significantly higher redshifts and star-formation rates
than the radio-selected objects in this paper: $\langle z \rangle \sim
2.5$ and $\langle L_\mathrm{FIR} \rangle \sim 4\times
10^{12}{\rm L}_{\odot}$ for the submm-selected galaxies, versus $\langle z
\rangle \sim 1$ and $\langle L_\mathrm{FIR} \rangle \sim 1\times
10^{11}{\rm L}_{\odot}$ for the faint radio sources discussed here.  Thus the
size differences are broadly consistent with an extrapolation of the
$r_\mathrm{e}$--$L_\mathrm{FIR}$ relation of \cite{Fujimoto17} to
lower FIR luminosities.  

\section{Summary}
A catalog of 792 radio sources at 3~GHz was derived from sensitive
($\sigma \approx 1~\mu\mathrm{Jy~beam}^{-1}$) VLA
S-band ($2 <\nu < 4 \mathrm{~GHz}$)
images having $3\arcsec$ and $0\,\farcs66$ resolution.
The reliability of the source detections was evaluated and is supported
by identifications with IR sources from the {\it Spitzer}
observations described in \citet{Mauduit2012}. 
Because 97.7\% of the radio sources have IR counterparts,
the reliability of sources in the catalog should be $\gtrsim 98$\%.
The strong FIR/radio correlation between the  flux densities
the radio detections and  their far IR counterparts shown in
Figure~\ref{ArcadeHerschel160} suggests that the radio sources are
dominated by star formation.
Polarized radio emission was detected in only six source components.


The typical source size was estimated from the statistics of the ratio
$S_\mathrm{p}^\mathrm{H} / S_\mathrm{p}^\mathrm{L}$
of peak flux densities at $0\,\farcs66$ and 3$\arcsec$ resolutions.  92\%
of the sources detected at the lower resolution were also detected in the
high-resolution image.
Most of the nondetections are at the
lowest flux densities, hence likely due to resolution and limited brightness
sensitivity.  Both Figures \ref{RatioVFlux} and \ref{RatioAOnly}, as
well as Table~\ref {SizeTab}, show that at peak sky flux densities
high enough that the distribution is not truncated by sensitivity
($\sim 20\, \mu\mathrm{Jy~beam }^{-1}$) the distribution is centered
  on a ratio corresponding to a source with a circular Gaussian FWHM
   $\langle \phi \rangle = 0\,\farcs3 \pm 0\,\farcs 1$.  The scatter is larger than at corresponding flux
  densities of simulated point or $0\,\farcs5$ FWHM sources,
  indicating a range of intrinsic source sizes, with a few
  completely resolved at the higher resolution.  The rms scatter in
  the $S_\mathrm{p}^\mathrm{H} / S_\mathrm{p}^\mathrm{L}$ ratios of
  real sources is less than the difference in the ratios expected for
  unresolved and circular Gaussians of $0\,\farcs5$ FWHM; thus, the
  scatter about the typical size is fairly small.  Table~\ref{FakeTab}
  gives the analysis for the simulated sources and suggests that the
  sizes are slightly underestimated.  A simple analysis of the ratio
  statistics gives a distribution of the equivalent circular Gaussian
  FWHM, with a median $\langle \phi \rangle = 0\,\farcs3 \pm 0\,\farcs 1$
  and a population size scatter  $\sigma_\mathrm{pop} \lesssim 0\,\farcs3 $.

\acknowledgments
{\rev 
We thank the anonymous reviewer for the many comments and suggestions
leading to a clearer description of these results.
This work is based in part on observations made with the Spitzer Space
Telescope, which is operated by the Jet Propulsion Laboratory,
California Institute of Technology under a contract with NASA. 
This work is also based in part on observations made with Herschel, a
European Space Agency Cornerstone Mission with significant
participation by NASA. 
Herschel is an ESA space observatory with science instruments provided
by European-led Principal Investigator consortia and with important
participation from NASA.  
}



{\it Facility:} \facility{VLA}.

\end{document}